# The cloud paradigm:
## Are you tuned for the lyrics?

**Fernando Brito e Abreu**

DCTI / Instituto Universitário de Lisboa (ISCTE-IUL)
& QUASAR / CITI / FCT / Universidade Nova de Lisboa

*Abstract*

*Major players, business angels and opinion-makers are broadcasting beguiled lyrics on the most recent IT hype: your software should ascend to the clouds. There are many clouds and the stake is high. Distractedly, many of us became assiduous users of the cloud, but perhaps due to the legacy systems and legacy knowledge, IT professionals, mainly those many that work in business information systems for the long tail, are not as much plunged into producing cloud-based systems for their clients.*

*This keynote will delve into several aspects of this cloud paradigm, from more generic concerns regarding security and value for money, to more specific worries that reach software engineers in general. Do we need a different software development process? Are development techniques and tools mature enough? What about the role of open-source in the cloud? How do we assess the quality in cloud-based development? Please stay tuned for more!*

**Stairway to heaven** (Jimmy Page & Robert Plant, Led Zeppelin)

Cloud computing refers to the on-demand provision of computational resources (data and software) via the Internet, rather than from a local computer. Cloud users can submit a task, such as searching for a part in the company's inventory, to the service provider, without possessing the database and the stock management software in its premises. The client's computer may contain only a browser and a minimal operating system. Since the cloud is the underlying delivery mechanism, cloud-based applications and services may support any type of software application or service in use today. This on-demand delivery model is being increasingly used in many business applications, including accounting, collaboration, customer relationship management (CRM), enterprise resource planning (ERP), invoicing, human resource management (HRM), content management (CM) and service desk management.

One day I woke up and realized my data and software are indeed ascending to the clouds. Many of my important documents are in the *Dropbox*. My slides are on *Slideshare*. My written messages follow their way on *Gmail* and my spoken ones go through *Skype*. I keep in touch with my peers in *LinkedIn* and with my friends on *Facebook*. I schedule my meetings in *Doodle* and I do cooperative work with *GoogleDocs.* I teach my students to work in software engineering projects using *Sourceforge* or *GoogleCode*. In all cases these are cloud-based services. I hold a private account in all of them, and someone must be paying for it, since I am not. I am happy to possess the gift of being able to communicate with that remote and distributed soul of my computer, but sometimes I wonder where my data physically is and how will I manage if I lose contact with it.

Could computing data centers used by cloud service providers are usually not open for tours, but those providers tell us not to be like Thomas, the Jesus disciple, who somehow missed the first appearance of Jesus after his resurrection. We should have faith that our assets are well protected! However, it has been pointed out that most cloud providers do not even own and operate the platforms they provide you service on. Instead, they use shared infrastructures, so cross-pollination of service provisioning could affect portability, reliability and security. In doubt, I try to keep a copy of my critical stuff in a large (and cheap) storage at home, but that requires discipline.

Who will do that in a small to medium sized enterprise (SME) that bought the concept of cloud computing as a way of cutting costs with a local IT infrastructure, and therefore has no dedicated operation staff?

## Money (Roger Waters, Pink Floyd)

As a private user I (still) do not pay for those aforementioned cloud services. What will I do if their providers decide to suspend this "free" lunch? At this point I will hear the unison scream of all individual cloud users: we are already paying the bill by being bombarded with web advertising associated with those cloud services! And we are asked to pay if we want to get beyond the basic service. OK, it is fair, so let us have faith that the "basic" service that hooked millions of us on those cloud services will keep being provided for free. I cannot live without some essential commodities such as water, electricity, gas, insurance, cleaning, tv and web assess, but I would not be willing to pay 10 extra bills, one for each of the aforementioned cloud services I am currently using. However, we must be aware that, since it is easy to control the access to cloud-based applications, some major players in the horizontal arena (general-purpose programs addressing the needs of many people) see the cloud as an opportunity to get rid of their haunting nightmare of software piracy.

There is another economic concern that assaults my spirit. Usually we do investments that are supposed to pay themselves in the medium to the long run. That is what we expect when buying real estate, or when a farmer buys seeds and fertilizer, or when the owner of a factory purchases a new machine for his plant, or even when we pay the fees of a good school for our children. All of these ventures seek to reach the break-even point, the point in time where the return on investment (ROI) becomes positive. That logic is the same for those big players that in recent years have invested multi-billion dollars in setting up impressive data centers (e.g. Amazon, Google, Microsoft or Salesforce). Who will guarantee their ROI? Big companies (e.g. telcos, banks, insurance, big wholesalers) will have their privately held clouds, both because they have the know-how and can afford it, and because they are very much concerned with securing their assets. SMEs are much more prone to buy the concept of cloud computing and they represent a large proportion of the whole market in most countries nowadays. This is the basis of the Long Tail theory. However, note that the ROI rationale is sold upside down by cloud providers. Pay attention to NIST[1] definition of cloud computing:

*"Cloud computing is a model for enabling convenient, on-demand network access to a shared pool of configurable computing resources (e.g., networks, servers, storage, applications, and services) that can be rapidly provisioned and released with minimal management effort or service provider interaction."*

The message forwarded to SME administrators is the following: cloud computing customers do not own the physical infrastructure, instead avoiding capital expenditure by renting usage from a third-party provider. They consume resources as a service and pay only for resources that they use. Cloud computing users can avoid capital expenditure (CapEx) on hardware, software, and services when they pay a provider only for what they use. Consumption is usually billed on a utility (resources consumed, like electricity) or subscription (time-based, like a newspaper) basis, with little or no upfront cost. The idea of not having to make an initial investment is tempting for our SME executives. Paying a variable amount depending on the number of transactions seems fair, because more transactions mean more business. Sometimes they forget that the volume of data stored will be an ever-increasing portion of their cloud bill and will not decrease when business will slow down. Perhaps more worrying should be the argument of the minimal bootstrap effort. What is the cost of reverting that model to a cloud-computing based approach? One obvious cost will be that of migrating legacy databases required to keep doing business as usual. And what about migrating to the cloud the customized applications that are often found in SMEs?

---

[1] - National Institute of Standards and Technology, an American agency in the Technology Administration that makes measurements and sets standards as needed by industry or government programs.

## The times they are a-changin' (Bob Dylan)

A pervasive thought of all SME stakeholders is increasing productivity and that can be achieved if they cut on labor costs. Most SMEs invested in their computerization in the latter quarter of the twentieth century. They have some IT staff or external consultants that guarantee the smooth operation and the evolution of their applications, when requirements change. Cloud computing sounds as a great opportunity to get rid of those costs and attain a better ROI. Globally, the consequences of a widespread adoption of cloud computing in the IT labor market for SMEs may seem devastating. Moreover, we will not need to train as many software engineers as we do today. Maybe I even face the risk of getting out of business as a computer science professor. I will share with you some reasons why we should not be scared with this scenario. However, some action will be required, or somebody will take the cheese away from you.

The first thing we must rationalize is where in the cloud puzzle we can play our role as an IT professional. Cloud services are provided at three levels of abstraction: infrastructure, platform and software. Let us have a look at each one of them separately.

*Infrastructure-as-a-Service (IaaS)* stands for delivering a platform virtualization environment as a service, along with scalable raw storage and networking capabilities. Rather than purchasing servers, data-center space or network equipment, clients buy those resources as a fully outsourced service. The suppliers typically bill such services on a utility computing basis. The amount of resources consumed (and therefore the cost) will typically reflect the level of activity.

*Platform-as-a-service (PaaS)* is the most important level of abstraction for software developers in the cloud since it encompasses the provision of all facilities required to support the complete life cycle of building and delivering cloud applications / services. Those facilities include tools for application design, development, testing and deployment, web service integration and marshalling, database integration (providing persistence, transactions, security and concurrency capabilities), application versioning, issue tracking and team collaboration. The orchestration and integration of these services is of utmost importance, since that is how we define and deploy a software process.

*Software-as-a-Service (SaaS)* is the on-demand software delivery model in which software and its associated data are hosted centrally (in the cloud) and are typically accessed by users using a thin client (usually a web browser). SaaS corresponds to the business / users perspective that we have already tackled in a previous section, so we will not go any further on this.

In retrospect, IaaS is mostly a matter for systems engineers and networks administrators, while SaaS is largely an issue for users, marketing staff and business administration staff. Let us then concentrate on what software engineers should care the most. PaaS enables you to create cloud applications, without the cost and complexity of buying and managing the underlying software/hardware. In my talk I will briefly overview the following mainstream PaaS platforms:

- *Amazon Web Services (AWS)*
- *Google App Engine*
- *Microsoft Windows Azure*
- *Force.com platform*
- *IBM SmartCloud platform*

### Free as a bird (John Lennon, The Beatles) or Hotel California (The Eagles) ?

While the major players are interested in hooking you to their proprietary solutions, therefore guaranteeing that once you are in their cloud you will never get out, like in Hotel California, the open source community is fighting back with open solutions. One aspect of this open movement is the ability for companies to build their own "private clouds". A private cloud is an infrastructure operated solely for a single organization, whether managed internally or by a third-party and hosted internally or externally. Big cloud players are usually not pleased with this concept and the opinion-makers on their behest argue that with a private cloud, a user company will still have to buy, build, and manage it, therefore not benefiting from lower up-front capital costs and less hands-on management. Fortunately, this is not entirely true, due to some important ventures such as the [Eucalyptus](#) open-source project [2] that has developed an IaaS solution for private clouds that matches AWS API specifications. By September 2011 it had allowed to deploy more than 25,000 on-premise clouds.

On the PaaS side, there are very interesting open-source ventures, as well. In this keynote, I will briefly go through the following platforms:

- *VMware's [Cloud Foundry](#)*
- *WSO2's [StratosLive](#)*
- *RedHat's [OpenShift](#)*

### Good is Good (Sheryl Crow) or Bad (Michael Jackson) ?

We are used to assess software quality according to two perspectives: product quality and process quality.

The bottom-line will always be striving for good software products. Several software quality models have been proposed in the literature and industry has reached some consensus with the ISO/IEC 9126 standard that defines software quality characteristics, both on the users' and developers' viewpoints.

Taking the users' viewpoint, the ISO/IEC 9126 Software Quality in Use model proposes four characteristics: *Effectiveness, Productivity, Safety* and *Satisfaction*. In this keynote, I will go through each of those characteristics to conclude that there are few noticeable distinctions when comparing to non-cloud-based software products.

Regarding developers' viewpoint, the ISO/IEC 9126 standard defines a quality model composed of six quality characteristics: *Functionality, Reliability, Usability, Efficiency, Maintainability and Portability*. Each characteristic is subdivided into sub-characteristics and further described by using appropriate external and internal quality attributes that can be measured by using specified metrics. I will go through each of the aforementioned characteristics and try to identify the most distinctive ones for cloud-based software. Will we need to extend this set of characteristics to encompass specific aspects of cloud computing applications?

Last, but not the least, we get to the other software quality perspective: process quality! Product quality cannot be achieved consistently without a well-defined process. Several process quality assessment and improvement frameworks, such as SPICE, CMMI or the MPS.Br, have been proposed. These frameworks rely on the identification of best practices on software development resulting from lessons learnt for more than two decades. Before assessing the cloud development process, we have to discuss if the "old" best practices still apply and/or identify other relevant practices that are specific to the cloud-based software development lifecycle. I believe this is where Software Engineering researchers still need to dig a lot, mainly by launching action research endeavors in cloud-based projects.

---

[2] - at the University of California at Santa Barbara, home of 5 Nobel laureates